\documentclass[conference]{IEEEtran}
\IEEEoverridecommandlockouts
\usepackage{cite}

\usepackage{multirow}

\usepackage{algorithmic}
\usepackage[linesnumbered,ruled,vlined]{algorithm2e}
\usepackage{graphicx}
\usepackage{textcomp}
\usepackage{xcolor}
\usepackage[ruled]{algorithm2e}
\usepackage[stable]{footmisc}
\usepackage{balance}

\def\BibTeX{{\rm B\kern-.05em{\sc i\kern-.025em b}\kern-.08em
    T\kern-.1667em\lower.7ex\hbox{E}\kern-.125emX}}
\begin{document}

\title{Optimizing Tail Latency in Commodity Datacenters using Forward Error Correction \\
}

\author{\IEEEauthorblockN{Gaoxiong~Zeng$^1$,~~~~~Li~Chen$^2$,~~~~~ Bairen~Yi$^3$,~~~~~Kai~Chen$^1$} \IEEEauthorblockA{\textit{$^1$iSING Lab@Hong Kong University of Science and Technology~~~$^2$Huawei~~~$^3$ByteDance}}}

\newcommand{\sys}{CloudBurst\xspace}

\maketitle
\thispagestyle{plain}
\pagestyle{plain}

\begin{abstract}
Long tail latency of short flows (or messages) greatly affects user-facing applications in datacenters. Prior solutions to the problem introduce significant implementation complexities, such as global state monitoring, complex network control, or non-trivial switch modifications. While promising superior performance, they are hard to implement in practice.

This paper presents \sys, a simple, effective yet readily deployable solution achieving similar or even better results without introducing the above complexities. At its core, \sys explores forward error correction (FEC) over multipath --- it proactively spreads FEC-coded packets generated from messages over multipath in parallel, and recovers them with the first few arriving ones. As a result, \sys is able to obliviously exploit underutilized paths, thus achieving low tail latency.
We have implemented \sys as a user-space library, and deployed it on a testbed with commodity switches. 
Our testbed and simulation experiments show the superior performance of \sys. For example, \sys achieves $63.69\%$ and $60.06\%$ reduction in 99th percentile message/flow completion time (FCT) compared to DCTCP and PIAS, respectively.
\end{abstract}

\section{Introduction}

Low latency message\footnote{Message and short flow are used interchangeably in this paper, both referring to small network flows.} delivery is critical for many applications in datacenter networks (DCN), such as web search~\cite{dctcp,pias-nsdi}, page creation~\cite{detail}, recommendation systems, stream processing, and online advertising~\cite{d2tcp}. These applications are usually user-facing, and even a very small delay in flow completion time (FCT) can reduce application performance, degrading user experience~\cite{dctcp,pias-nsdi,pias-hotnets} and causing financial loss~\cite{d2tcp}.

However, the long tail latency problem is particularly pronounced in production DCNs for multiple reasons ($\S$\ref{sec:background:cause}): (1)~applications emit synchronized high fan-in bursts (incast~\cite{incast, pac}); (2)~shared-buffer switches are too shallow~\cite{dc-buffer,dc-buffer-apnet} to absorb bursts; (3)~transport protocols use Automatic Repeat Request (ARQ)~\cite{gobackn,tcpsack} and retransmission timeouts for packet recovery; (4)~coarse-grained load balancing (e.g., ECMP~\cite{ecmp}); and (5)~hardware malfunctioning (e.g., packet black-hole, silent random packet drops, etc.~\cite{pingmesh,fuso,scalerdma}) is unpredictable.

In order to tackle such long tail latency problem, prior works ($\S$\ref{sec:background:prior}) adopt a variety of strategies from fine-grained load balancing~\cite{detail,conga,fastpass,hermes}, rate control~\cite{dctcp,d2tcp,dcqcn,timely,swift,hpcc,dcn-transport,mcp,mqecn,tcn,ecn-sharp}, prioritization~\cite{pias-nsdi,pfabric,pase,detail,qjump,ras,karuna}, to fast loss recovery~\cite{cp,fuso}. However, most of these proposals introduce non-trivial implementation difficulties, such as global state monitoring~\cite{conga,fastpass}, complex network control~\cite{fastpass,pase}, and switch  modifications~\cite{cp,pfabric,conga,hpcc}. While achieving superior performance, these solutions are hard to deploy in existing commodity datacenters.

Therefore, we ask a pragmatic question: \emph{is there a simple scheme for commodity datacenters to cut tail latency without the above complexities, while still delivering similar or better performance?} In this paper, we answer the question affirmatively by presenting \sys, a simple and deployable solution to cut tail latency of short messages in datacenters.

\begin{figure}[t]
  \centering
  \vspace{-1.2em}
  \includegraphics[width=1\linewidth]{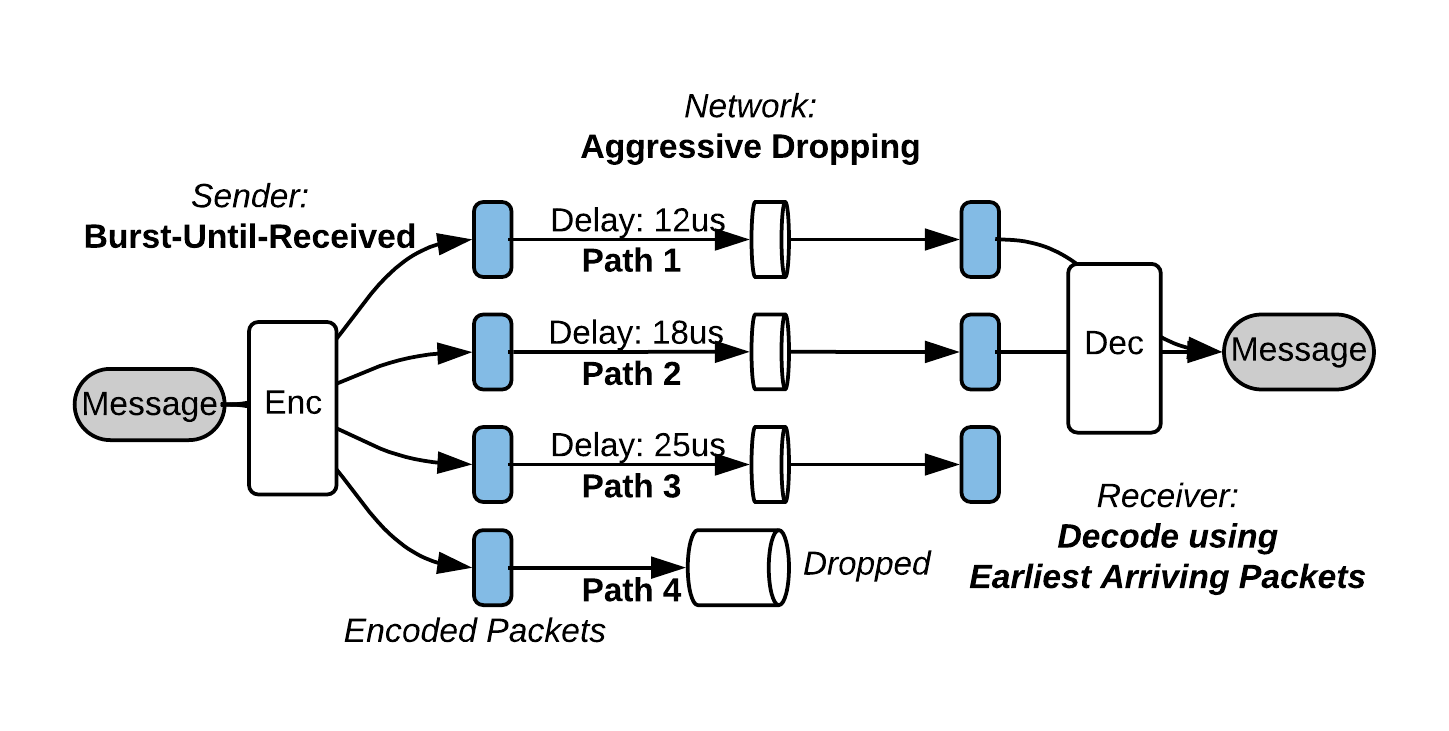}\\
  \vspace{-2em}
  \caption{\sys Concept}\label{fig:sys}
  \vspace{-1em}
\end{figure}

\sys works as follows (Figure~\ref{fig:sys} \& $\S$\ref{sec:design}):

\begin{itemize}
\item \textit{Encoding messages with FEC:} Forward error correction (FEC) has been deployed in many applications~\cite{ffc, oec}, it uses proactive and redundant approach to tolerate errors. In DCN, the error tolerant feature can be adopted to handle packet losses. That is, we do not need to retransmit the lost packets while still guaranteeing the reliable transmission. 
\sys explores the coding dimension of the transport-layer design. It employs FEC with proactive and oblivious redundant transmission. Each encoded packet contains information of multiple original packets. In this way, even if some of the encoded packets are lost, the original packets can still be reconstructed at the receiver.

\item \textit{Bursting over multiple paths:} \sys spreads encoded packets over multiple paths, which obliviously exploits the rich path diversity in modern DCNs~\cite{fattree, facebook-dcn, google-dcn}, as well as the temporary network under-utilization~\cite{dctraffic10}. If any congestion-free path exists, \sys will take advantage of them without extra signalling overhead.

\item \textit{Separated and size-limited switch queue:} We limit the buffer usage of the \sys short flows to minimum, so that the per-hop queueing latency of a \sys packet is deterministically low (less than a few $\mu s$). This is achieved by separating \sys short flows and other traffic in different queues, and limiting the maximum depth of the \sys queue to a tiny value (a few packets). Such a limitation on buffer usage of \sys traffic also makes it friendly to other traffic.
\end{itemize}

Our design choices result in a very simple transport that works drastically different from TCP. \sys performs neither congestion control\footnote{In fact, for erasure coded flows, dropping can be considered as a form of congestion control~\cite{geni}, or ``decongestion control''~\cite{decongestion}.} nor reactive retransmission (i.e., no per-packet ACK, no congestion feedback, no loss detection, and no retransmission timeout (RTO)).
Instead, it simply transmits at line-rate and unilaterally keeps generating and sending the encoded packets until the receiver acknowledges the reception of the message. Meanwhile, the in-network requirement of \sys is also simple: limiting queue size is readily configurable in commodity switches\cite{pica8}.

To be deployable with existing commodity datacenters, \sys is now implemented as a user-space library ($\S$\ref{sec:implement}), so that applications can voluntarily use it for latency-sensitive short flows.
We built a small 2-tier leaf-spine testbed and deployed \sys on it. The testbed consists of 6 Pronto-3295 (Broadcom) Ethernet switches and 8 Dell R320 servers with quad core Xeons E5-1410 CPU and 1GbE NIC. Our implementation experience shows that \sys is readily-deployable with existing commodity switches.

Our testbed experiments ($\S$\ref{sec:basicdesign}--$\S$\ref{sec:deepdive}) show that \sys outperforms prior practical schemes. For example, it achieves more than $60.06\%$ and $63.69\%$ reduction in 99th percentile (p99) flow completion time compared to PIAS~\cite{pias-nsdi} and DCTCP~\cite{dctcp}. 
Furthermore, our results also verify that it is resilient to incast, 
and handles failures gracefully.
We complement testbed experiments with large-scale simulations ($\S$\ref{sec:simulations}) in 10/40G environments. We find that \sys achieves comparable or better performance than prior solutions. For example, it achieves $24\%$ tail latency reduction compared to a clean-slate design, pFabric~\cite{pfabric}.


\section{Achieving Low Latency}\label{sec:background}
We first look at the causes for long tail latency in DCN ($\S$\ref{sec:background:cause}). Then, we review the prior solutions to cut tail latency ($\S$\ref{sec:background:prior}). Finally, we overview \sys ($\S$\ref{sec:background:lowl}).

\subsection{Causes of long tail latency}
\label{sec:background:cause}

The traffic characteristics of DC applications, current DCN configurations, and the usage of TCP in DCNs jointly contribute to the long tail latency for short flows.

\textbf{High fan-in bursts (incast~\cite{incast}):} Tree-based applications emit synchronized high fan-in bursts. They are constructed with multiple layers, and each parent aggregates results from its children. The interactivity constraint (e.g. 300ms\cite{d2tcp}) is distributed to each layer (e.g. 40ms). Hence children of a same parent respond at almost the same time, causing fan-in burst at the parent. These synchronized fan-in bursts exceed the egress port capacity of switches, causing queue to build up and may lead to congestive packet drops.

\textbf{Shallow shared-buffer switch:} Shared buffer is usually configured to absorb bursts, i.e. a portion of buffer is shared among multiple ports, so if one port is experiencing bursts, it can take all available shared buffer. However, shallow buffer switches cannot absorb fan-in bursts from applications at high bandwidth. The expansion of buffer (from 4MB to 12MB~\cite{dc-buffer}) in commodity switches is not compatible with the 10X$+$ growth of per port bandwidth (from 1Gbps to 10/40Gbps~\cite{google-dcn, gemini}). Additionally, throughput-intensive flows requires a certain amount of buffering
($\alpha \times$Bandwidth Delay Product, $\alpha$ depends on the transport protocol~\cite{dctcp}). in the switch to achieve high throughput, requiring guaranteed buffer in each port and further reducing the shared buffer for burst tolerance. When bursts occurs, packets from latency-sensitive flows may be dropped due to lack of buffer~\cite{incast}.

\textbf{Error handling and retransmission timeout:} Applications deliver data reliably using TCP with Automatic Retransmission Request (ARQ) error handling (e.g., Go-back-N~\cite{gobackn}, Selective Repeat~\cite{tcpsack}), and timeout-based error discovery). For each error, ARQ needs at least one RTT to recover, thus it works well for long flows, because acknowledgement of received or missed packets are batched for large congestion windows. However, ARQ does not help for short flows, which may finish within the slow start phase. If a flow's first few packets coincide with congestion and get dropped, it can only discover the loss by RTO. Therefore, setting a proper RTO$_{min}$~\cite{detail,incast} is key to fast recovery of short flows. RTO$_{min}$ in DCN is usually set to 5ms~\cite{incast,fuso}, which is almost two orders of magnitude larger than the base RTT ($\sim100$us). A single timeout can easily lead to long latency tail. Reducing the timeout can surely benefit packet recovery, but may also increase the load on network and servers due to frequent retransmissions and require high-precision timers.

\textbf{Malfunctioning hardware:} Packet loss may also occur due to hardware failures, which happens even in well-engineered modern DCNs with lossless fabric~\cite{fuso,scalerdma,pingmesh}. Such failures may come from TCAM deficits, aging transceivers, etc. Particularly, silent packet drops or blackholes are discovered with network-wide diagnostic tools~\cite{pingmesh}. Thus, these failures are difficult for single-path transport to recover, and can take multiple RTTs for multipath transport to recover.

\textbf{Imperfect flow load balancing (LB):} Current load balancing in commodity DCNs usually depends on flow hashing, i.e., per-flow ECMP~\cite{ecmp}, to keep utilization of parallel paths even. However, hash collision may occur, which can temporarily overload some links and cause queue length to grow, prolonging per-packet latency and inducing packets drops. Randomized LB also cannot help with application bursts when all flows have the same destination port.

\subsection{Prior Solutions}\label{sec:background:prior}

We overview the prior solutions to the tail latency problem: 

\textbf{Reducing queueing latency:} Generally, solutions in this category follow a few strategies such as fine-grained load balancing~\cite{conga,fastpass,detail,hermes}, rate control~\cite{dctcp,d2tcp,hpcc,dcqcn,timely,swift,dcn-transport,mcp,mqecn,tcn,ecn-sharp}, and traffic prioritization~\cite{pfabric,pias-nsdi,pase,detail,qjump,ras,karuna}.
\begin{itemize}
\item By load balancing traffic across multiple paths evenly, we can avoid excessive queue build-up that may result in queueing latency or loss. For better performance, congestion awareness is often required~\cite{conga,detail,fastpass,hermes}. For example, CONGA~\cite{conga} measures link utilization and directs flows to less congested paths, while DeTail~\cite{detail} detects and avoids congestion by monitoring queue lengths. In the extreme, Fastpass~\cite{fastpass} centrally schedules and routes every packet with complete knowledge of per-path load conditions. These solutions, however, require very complex network control or switch modifications.

\item By rate-throttling flows (especially the larger ones) based on ECN~\cite{ecn}, delay~\cite{ecn-rtt}, or in-band network telemetry (INT)~\cite{hpcc} signal, we can control the queue build-up so that latency-sensitive short flows see small queues at the switch.
While helpful, these solutions~\cite{dctcp, d2tcp, dcqcn, timely, hpcc, swift} still rely on load-balancing, and short flows may suffer long latency when traffic is unevenly distributed. Besides, the advanced INT signal may not even be available on switches and thus require customized hardware.

 \item By giving latency-sensitive flows high priority, the switch dequeues them first, without regards to lower priority ones before them, thus achieving low latency~\cite{pias-nsdi, pfabric, pase, detail, qjump}. For example, pFabric~\cite{pfabric} assumes priority dequeueing (and dropping) to minimize flow completion time of short flows, and QJump~\cite{qjump} leverages prioritization to cut tail latency. However, they either assume infinite switch queues or rely on accurate configurations.
\end{itemize}

\textbf{Recovering from packet losses:} Fast in-network feedback for packet drops~\cite{cp,ndp} can accelerate the transition of TCP state machine, thus triggering retransmission earlier. As above, these solutions also require non-trivial hardware modifications. Retransmission can also be proactive: FUSO~\cite{fuso} augments MPTCP~\cite{mptcp,mptcp-dcn} by eagerly retransmitting on less congested paths. However, it needs to keep complex states of subflows.

Another line of work~\cite{dcqcn,timely,pcn} seeks help from lossless fabric~\cite{pfc}. However, at large scale, packet loss may still happen even on lossless fabric, due to mis-configuration or hardware failures~\cite{scalerdma}. In case of losses, they often rely on the NIC hardware for efficient recovery. This line of work is still under exploration and is beyond the scope of this paper.

\textbf{Proactive transport solutions:} Some recent works~\cite{expresspass,homa,ndp} use pre-allocated rate to avoid excessive packet delay and inaccurate self-adapted rate control. In these solutions, link capacities are {\em proactively} allocated by the receivers as ``credits'' to each active sender who then send ``scheduled packets'' at an optimal rate to ensure low queueing delay and near-zero packet loss. However, this approach requires at least one RTT to allocate credits to a new flow, which is unacceptable for short flows.
While recent augmentation scheme~\cite{aeolus} enables line rate start, issues still exist for short flows. For example, it is hard to assign a right amount of credits to senders before termination in the last RTT~\cite{flashpass}---too large for short flows will lead to link under-utilization due to credit wastage, while too small will introduce large delay.


\subsection{Cutting tail latency via FEC over multipath (Overview)}\label{sec:background:lowl}
\sys aims to cut the long tail latency with a simple and readily deployable protocol. It applies FEC to multiple paths ($\S$\ref{sec:fec}) in commodity datacenters. By proactively spreading encoded packets over multiple paths in parallel, and decoding the first few arriving ones to recover the original message, \sys obliviously exploits the uncongested paths to achieve persistent low latency, while not maintaining network state or performing complex control.

At the end-host, \sys performs ``burst-until-received'' ($\S$\ref{sec:protocol}). The sender encodes short messages with FEC and proactively sends the encoded packets over multipaths until the messages are decoded at the receiver with the first few arriving packets. In this way, \sys always exploits the best paths for low latency. If any uncongested path exists, \sys uses it without signaling; if a path is experiencing congestion or failure, \sys avoids it automatically by using encoded packets from other under-utilized paths.

At the switch, \sys performs aggressive dropping ($\S$\ref{sec:aggdrop}) for \sys flows, We limit the buffer usage of the \sys flows to minimum, i.e., separating \sys traffic and other traffic in different queues, and limiting the maximum depth of the \sys queue to a small value. This ensures that the per-hop queueing latency is deterministically low. Besides, it protects other flows from being affected by the burst-until-received behavior of \sys flows, making it friendly to co-existing traffic.

\section{Design}\label{sec:design}
We proceed to describe the details of \sys design.

\subsection{Choosing FEC for encoding}\label{sec:fec}
\begin{figure}[t]
  \centering
  \vspace{-1em}
  \includegraphics[width=\linewidth]{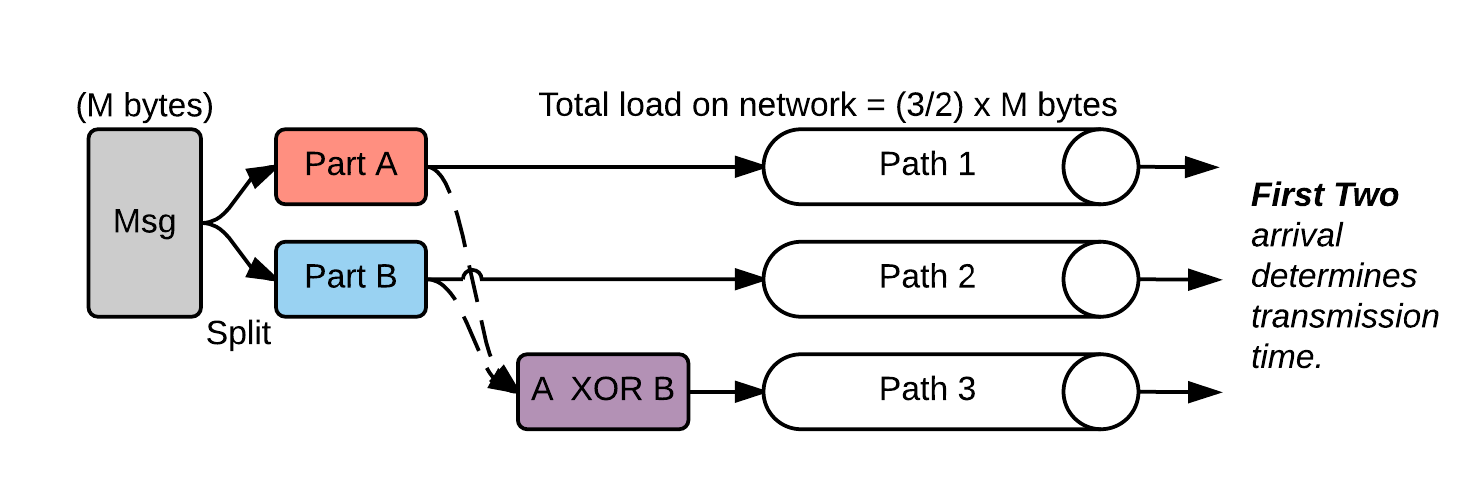}\\
  \vspace{-1em}
  \caption{Example of a 3-path FEC}\label{fig:blockcode}
  \vspace{-1.5em}
\end{figure}

Figure~\ref{fig:blockcode} is a simplified illustration of how FEC works. There are 3 paths from source to destination, and past experience indicates only one of them has congestion at any given time, but the sender does not know which until it sends.
Suppose the sender has 2 packets to send, say $A$ and $B$. A suitable FEC scheme is to create another packet $C=A\otimes B$ (bitwise exclusive-OR). After encoding, the sender sends $A$, $B$, $C$ on three paths in parallel. The receiver can recover the message as soon as it collects any 2 out of 3 packets, thus avoiding congestion at the cost of sending 1.5x more traffic.

Fixed-rate block codes, as shown in the example in Figure~\ref{fig:blockcode}, are simple to understand and implement, but have the fundamental difficulty of choosing a right coding rate.
In Figure~\ref{fig:blockcode}, the suitable code rate is $\frac{3}{2}$. Fixed rate codes like this may suffer if the path condition deteriorates: if two of the paths suddenly become congested instead of one in Figure~\ref{fig:blockcode}, then the message will take much longer time to recover, as the first two arrivals determine the transmission time.

Rateless erasure codes, or fountain codes~\cite{rateless}, are better suited for dynamic traffic characteristics in DCNs, as its code rate is adaptable to dynamic channel conditions.
The key property of fountain codes is that they can generate a potentially limitless sequence of encoded symbols from a given set of source symbols (the source symbols can be decoded with a large enough subset of encoded symbols).
Specifically, we adopt LT Code (LTC)~\cite{ltc} in our prototype due to its low complexity, where an encoded packet is generated by random linear combination of original packets.

\begin{figure}[t]
  \centering
  \includegraphics[width=\linewidth]{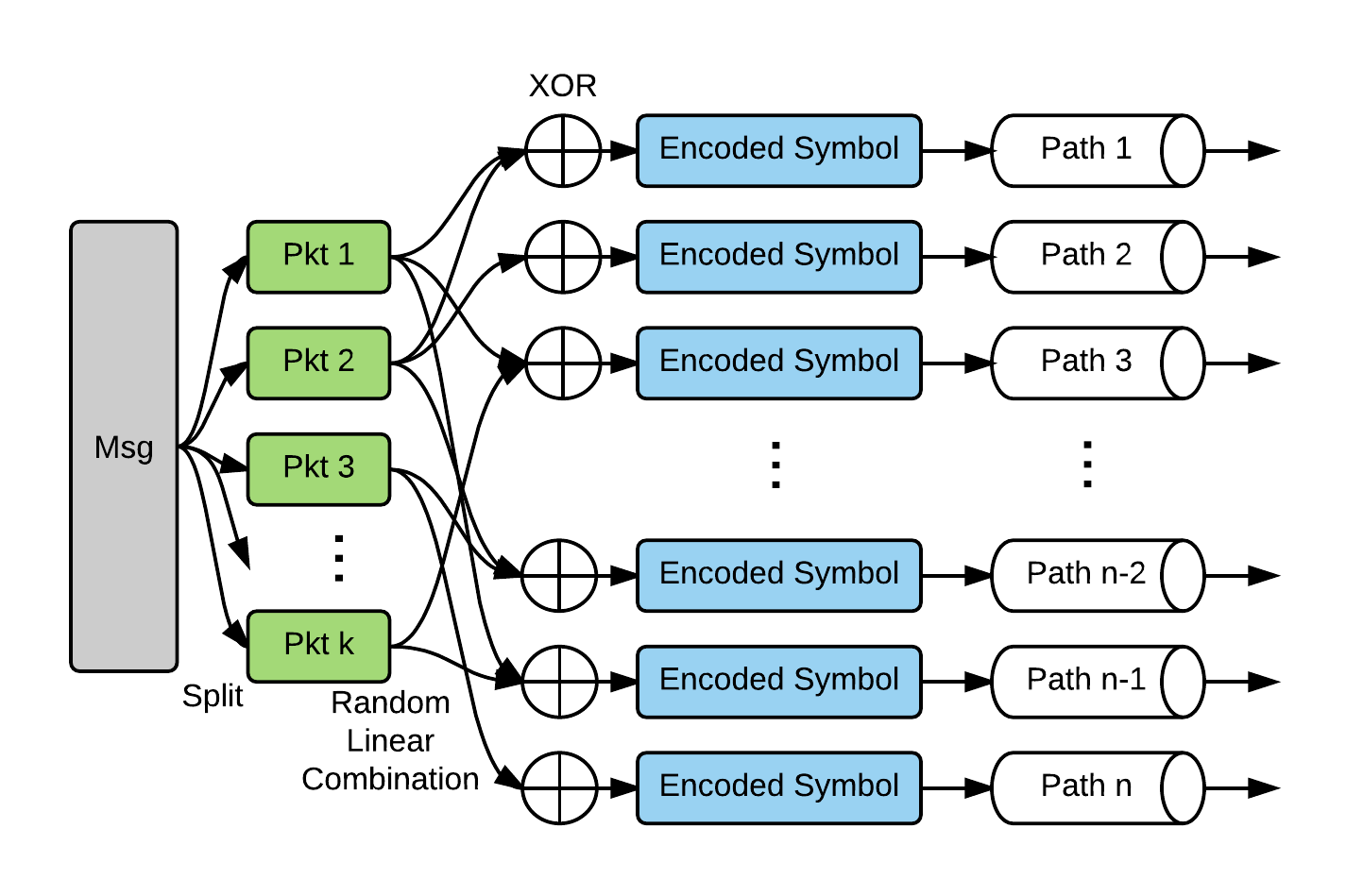}\\
  \vspace{-1em}
  \caption{Random linear encoding in one round of transmission}\label{fig:fountain}

  \vspace{-0.5em}
  \includegraphics[width=\linewidth]{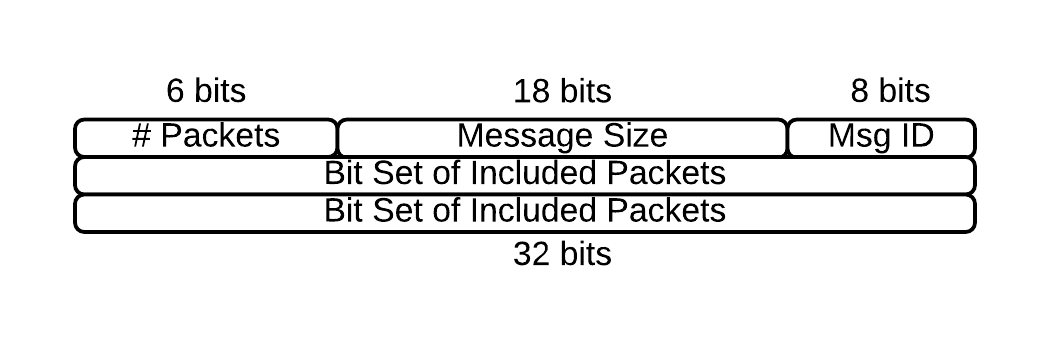}\\
  \vspace{-2.5em}
  \caption{\sys header format}\label{fig:header}

  \vspace{0.5em}
  \includegraphics[width=\linewidth]{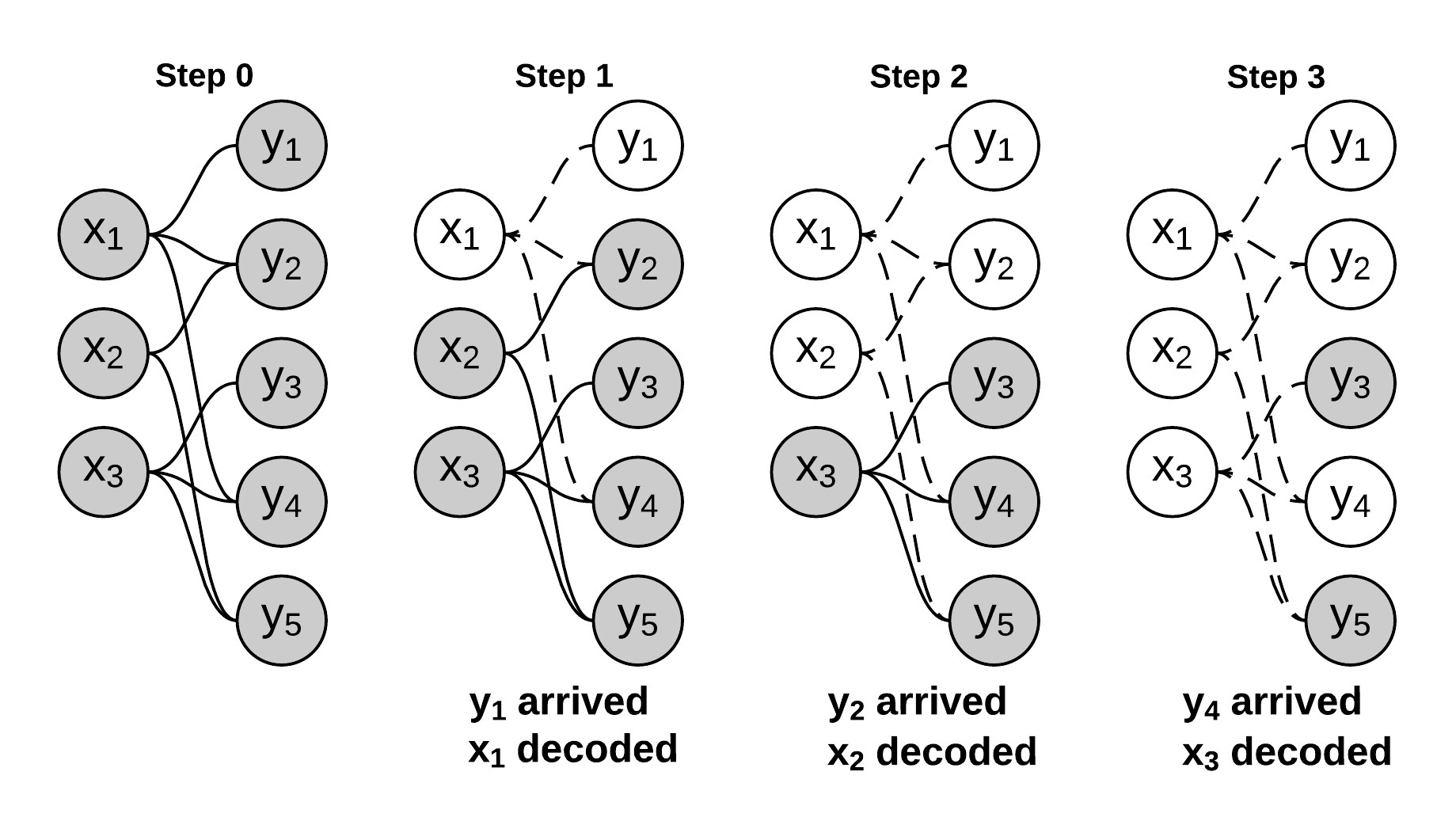}\\
  \vspace{-1em}
  \caption{Decoding example}\label{fig:dec}
  \vspace{-1em}
\end{figure}

\textbf{Encoding:}
We assume the message is given before the transmission.
As shown in Figure~\ref{fig:fountain}, we first packetize the original message into equal-length parts.
If necessary, we pad zero bits to packets so that every part from the original message has the same size: $MTU-H_{IP}-H_{UDP}-H_{cbrst}$, where MTU is Maximum Transmission Unit (1.5KB), $H_{IP}$ is IP header size (20B), $H_{UDP}$ is UDP header size (8B), and $H_{cbrst}$ is \sys header size.
For each encoded packet, its degree ($d$) is defined as the number of un-decoded original packets encoded in it. We first chose the value of $d$ from a certain probability distribution. Then, we randomly choose $d$ packets from the original message $\{x^{(1)},\ldots,$ $x^{(d)}\}$. The encoded packet in the $i$th iteration is then:
$y_i = x_i^{(1)}\oplus x_i^{(2)} \oplus \ldots \oplus x_i^{(d)}$.
The header of encoded packet (Figure~\ref{fig:header}) includes the number of packets ($n$) in the message, the message size in unit of bytes, an $8$-bit message ID, and the list of indices $(1), \ldots ,(d)$.

\textbf{Decoding:} The decoding algorithm uses standard Gaussian elimination method~\cite{rfc6330}. 
We illustrate the decoding with an example in Figure~\ref{fig:dec}. In step 1, $y_1$ is received, and its degree is 1 (Degree of a symbol is defined as the number of un-decoded original packets encoded in it), so that the corresponding $x_1$ is decoded. In step 2, $y_2$ is received, and its degree is 2. But since $x_1$ is already decoded, $y_2$'s degree is reduced by 1, so $x_2$ is decoded. By iteratively finding symbols with degree 1, the original message is recovered.

\subsection{Burst-until-Received at End-host}\label{sec:protocol}
\begin{figure}[t]
  \centering
  \vspace{1em}
  \includegraphics[width=\linewidth]{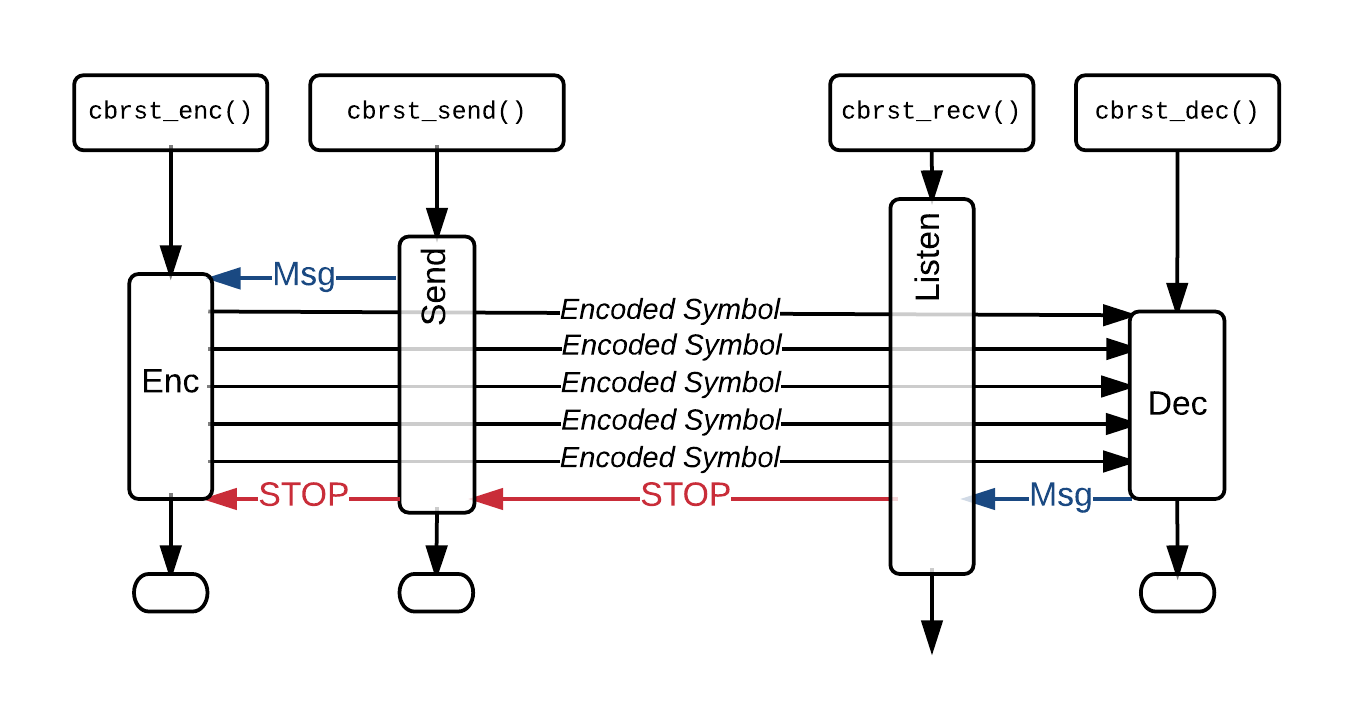}\\
  \vspace{-2em}
  \caption{Sequence Diagram}\label{fig:timeseq}
  \vspace{-1em}
\end{figure}

The sequence diagram of the end-host operations of \sys is shown in Figure~\ref{fig:timeseq}. We elaborate on the sender and receiver operations below.

\textbf{Sender operations:} As a variant of fountain code, LT coding can generate endless sequence of encoded packets for a given message. LT coding enables the \sys sender to continuously generate encoded packets and sending them on multiple paths, as described in Algorithm~\ref{algo:send}. It stops only after it receives a "STOP" signal from the receiver.

This burst-until-received behavior may consume all bandwidth of the network.
So we add a rate control mechanism to allow users to adjust the sending rate. The user/application specifies rate $r$, which is the share for \sys traffic. We define a round of transmission as the period of time the sender puts an encoded packet on each of the paths (Line\#6-9 in Algorithm~\ref{algo:send}), and the rate is adjusted by adding delay between rounds of transmission.
For example, if NIC capacity is 10Gbps, $r=0.5$, and 5 paths between source and destination. Then the sending rate of the message is 5Gbps (1Gbps per path). Pumping 5 1.5KB encoded packets on 5 paths takes 12$\mu$s, thus, a 12$\mu$s inter-round delay is added.

\begin{algorithm}[t]
\scriptsize
\DontPrintSemicolon  
\KwIn{Receiver IP \& Port, Message $m$, Transmission timeout $t$, Ratio $r$}
\KwOut{Success/Failure}
\uIf{sizeof($m$) $>$ MAX\_MSG\_SIZE}{
    \Return{Failure}\;
}
Set up count down timer of $t$\;
\While{STOP not received}{
  Wait for ($\frac{1}{r}$ * NUM\_PATHS * PKT\_SIZE / LINK\_CAP)\;
  \ForEach{path $p$ to Receiver}{
      Get symbol $y$ from encoding thread and send it on $p$\;
      \uIf{timer expires}{
      	\Return{Failure}\;
      }
  }
}
\Return{Success}
\caption{\texttt{cbrst\_send($\cdot$)}}
\label{algo:send}
\end{algorithm}

\textbf{Receiver operations:}
All encoded packets are received by the listening thread, and forwarded to their corresponding decoder (each decoder is instantiated for a message).
Decoder buffers all the encoded packets that have unrecovered packets for decoding. It returns the message to the receiver listening thread after full recovery, or expires if there is no more incoming encoded packets after a pre-defined timeout. Then the receiver signals a STOP to the sender. In case that the STOP signal is lost, the receiver will send a new STOP for each received encoded packet after the first STOP is sent.


\subsection{Aggressive Dropping in Network}\label{sec:aggdrop}
A major concern is that the open-loop rate control of \sys is dangerous to other flows and may overflow switch buffers in the network. Thus, it seems that a congestion control mechanism for \sys is needed.
However, dropping is also a form of congestion control~\cite{decongestion} for high-bandwidth erasure coded flows. When packets are erasure encoded, dropping some is not an issue as the message can be recovered with a subset of the packets sent from the source.
Thus, for a network with purely erasure coded flows, the switches do not need deep buffers to keep network stable~\cite{nocc}, and the sources can burst as fast as possible~\cite{decongestion}.

This is also true for TCP-like transport. For example, pFabric~\cite{pfabric} features minimum TCP, a line rate transport, with shallow buffer and priority dropping at the switch. The flows with same priority achieves the max-min fairness despite dropping due to shallow buffer.
In fact, replacing congestion control with erasure coding has been discussed for future Internet~\cite{geni}, but there are fundamental difficulties. First, in public Internet, many transport protocols coexist, and bursts of a few erasure coded flows may take all the buffer and bandwidth, hurting others. Second, on the path of an erasure coded flow, if a packet gets dropped after it passes a few switches, the bandwidth it consumed is wasted. These ``dead packets''~\cite{decongestion} may induce congestion collapse: packet loss requires additional packets (more redundancy) to recover, which incurs more load in the network and more packet drops, forming a positive feedback loop. 

By leveraging the properties of DCNs, targeting only the short flows, and limiting buffer usage, we avoid the drawbacks of erasure coding transport.
\begin{itemize}
\item DCNs often have abundant multipath~\cite{fattree, google-dcn, facebook-dcn}. With encoded symbols on all paths, the network core with \sys is load balanced for short flows, and the only bottleneck is the egress switch. With just one bottleneck, ``dead packets'' are not likely to occur.
\item For long flows, using erasure coding is dangerous due to the positive feedback loop described above. However, \sys serves only short flows, which do not persist: because of the small sizes and the high bandwidth network, short flows dissipates quickly.
\item We enforce aggressive dropping by separating the buffer for \sys and other flows, and limiting the buffer usage of the \sys flows to minimum. This protects other flows from \sys traffic. Thus, fan-in bursts of \sys flows no longer affect others on a shared shallow-buffer switch; meanwhile, TCP flows can also safely use remaining buffer for high throughput.
\end{itemize}


\subsection{Discussion: two potential issues}\label{sec:design_discussion}


\textbf{Infiniteness of flow sending:} The termination condition of our algorithm is that the receiver can decode all of the original packets. However, since we do not use the ACK mechanism, the senders do not know which packets have already arrived, they just randomly chose packets, which may cause the failure of flow completion. In fact, the mathematical property of LTC~\cite{ltc} guarantees the flow completion. $k$ original packets can be recovered by $k+\mathcal{O}(\sqrt{k}\ln^2{k/\delta})$ encoding packets with probability of $1-\delta$. That means, we can bound the total number of encoding packets with a certain probability. Meanwhile, we can also set a packet sending upper bound to avoid the tail latency caused by unlimited sending. When a certain number of encoding packets are sent, the sender will request the receiver for the information of received packets.

\textbf{Aggravate last hop congestion:} In fact, our algorithm focuses on the tail latency caused by the different congestion condition of sub-paths. In DCN, the last hop may be the bottleneck and sending FEC-coded packets only aggravates the congestion. This problem is still caused by the agnostic of packets' receiving in senders. When the last hop congestion (e.g., incast) happens, even the slow flow may not be finished in one RTT, it gives us the opportunity and time to respond the packet receiving information to the senders. More specifically, when a flow does not finish within a certain time, the sender will ask the receiver for the arrival information. Even though our algorithm is not designed for solving incast problem, it eliminates the retransmission cost, thus the experiment result shows that \sys still performs well in incast scenario.
\section{Implementation}\label{sec:implement}
In this section, we describe implementation and parameter settings of \sys.
\subsection{Enabling \sys at End-host}\label{sec:implement:endhost}
For the end-host, we build a prototype \sys with Rust 1.6~\cite{rust} as a user-space library. The implementation is multi-threaded: sending (receiving) and encoding (decoding) are handled by different threads at the sender (receiver).
We discuss the settings of coding-related parameters as follows.

\textbf{Choosing message size $n$ \& degree $d$:}\label{sec:implement:size}
A larger header size supports larger message size at the cost of more header overhead and possibly higher load on the network (with the same code rate, a larger message generates more encoded symbols).
We consider header overhead (percentage of packet used for header), computation overhead (number of decoding operations), and coding overhead (number of encoded symbols to reconstruct the message) when choosing $n$ \& $d$.

Header overhead is as follows: $x$ is the length of the field representing the maximum number of packets in a message ($n=2^x$ packets). The message size is $log_2(MTU\times 2^x)=x+log_2(MTU)$ (bytes). The bit-set representing the packets included in the symbol is $2^x$ (bits). 
Thus, assuming 1.5KB MTU, the header overhead for $x=9$ is less than $7\%$, which suggests that, choosing $x<10$ is efficient.

\begin{figure}[t]
  \centering
  
  
  \includegraphics[width=\linewidth]{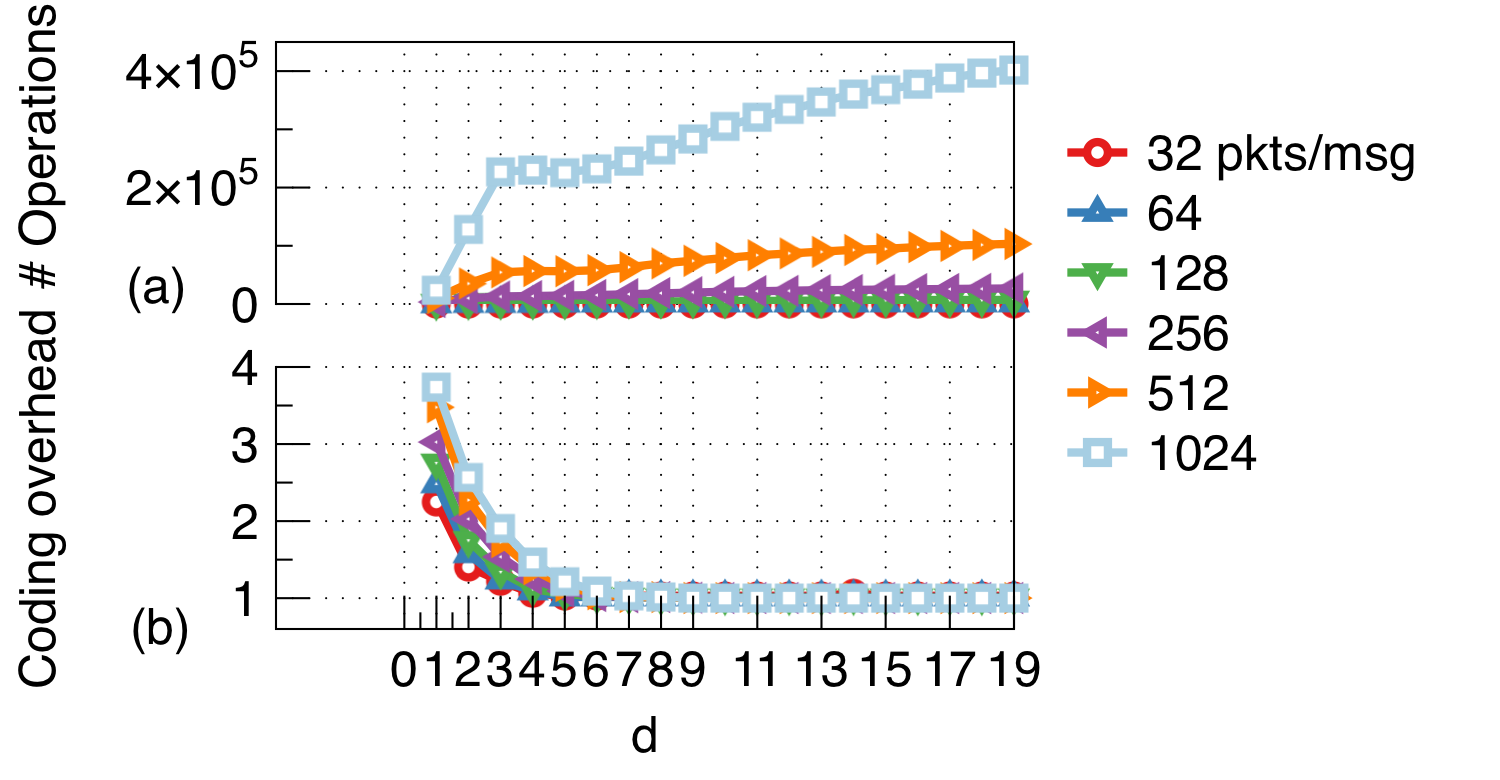}\\
  \vspace{-1em}
  \caption{Computation and coding overhead}\label{fig:coding-overhead}
  \vspace{-1em}
\end{figure}

Computation overhead for different encoding degree $d$ and message size $n$ is plotted in Figure~\ref{fig:coding-overhead}(a). For different $(d,n)$ pairs, we perform the en/decoding process 100 times, and calculate the average number of XOR operations necessary to reconstruct the message. We see that the number of operations increases with $n$. For each $n$, the number of operations increases with $d$, but slows down for $d\geq 5$.

Coding overhead is measured by the ratio between the number of packets needed to recovery the message and the number of original packets in the message. We collect this ratio by running the same experiment as above, and plot the results in Figure~\ref{fig:coding-overhead}(b). The coding overhead drops for all sizes with respect to $d$, but beyond $d\geq5$, increasing $d$ does not lead to lower coding overhead.

With the above results, we recommend to set $d=5, n=64=2^6$ for the current prototype. It thus supports message size of at most 93.44KB (i.e., $2^6 \times 1.46KB$).
We note that typical latency-sensitive applications in DCNs exhibit multi-tier partition/aggregation patterns, and the queries are often less than 10KB in size~\cite{dctcp,d2tcp}, which suggests the current header design should work well in practice.

\textbf{Choosing $r$:} $r$ is the expected throughput of \sys flows in the DCN. For oversubscribed DCN, we can adjust the parameter $r$ in Algorithm~\ref{algo:send} to the oversubscription ratio. For network with full bisection bandwidth, $r=1$.
In the experiments and simulations, we choose $r$ to be exactly the oversubscription ratio of the network, so that \sys flows will not overload the network.


\subsection{Enabling \sys in Network}\label{sec:impl:network}

\textbf{Configuring switch buffer:} To enforce aggressive dropping, we set the packets of \sys flows to the same DSCP value, so that they are carried in the same switch queue, separated from other non-\sys traffic. We then set the dedicated buffer for the \sys queue to a small value.

On our testbed, we set it to be $1$\% of total buffer size, and we disable the shared buffer of this queue to avoid affecting other flows. This is done by setting ``\texttt{buffer queue-limit}'' in our Pronto-3295 switches~\cite{pica8ref}. Other switches also support such configurations. For example, in Cisco switches, we can set the depth of a traffic class queue~\cite{ciscoref}.

\textbf{Multipath routing}
A key issue for \sys is how to spread packets on different paths. The implementation is dependent on how multipath is supported in the network.

\noindent\textit{Sub-optimal multipath:}
The network may use multipath implicitly, so that load balancing over multipath is transparent to the applications. ECMP~\cite{ecmp} is an good example: for each flow, an ECMP-enabled switch picks an outgoing port at random based on the hash of the flow's source and destination IPs and ports. To use \sys on ECMP, the sender and receiver need to maintain a pool of ports, and each encoded symbol will be given a header with a random combination of sender ports and receiver ports, which implicitly asks ECMP to hash packets on different paths. While this may not fully utilize all paths, we show that \sys still works well in $\S$\ref{sec:eval:multipath}.

\noindent\textit{Explicit multipath routing:}
Multiprotocol label switching~\cite{mpls} (MPLS) can provide explicit routing for each packet, but this requires support from the network fabric. Also, setting up multiple path labels for each short message requires signaling the switches on multipath, which is undesirable for low latency delivery.
To attain the same efficiency as implicit multipath with ECMP, we turn to a DCN routing scheme---XPath~\cite{xpath}, which enables explicit path-based routing in DCNs. It compresses and pre-installs end-to-end paths into forwarding tables of commodity switches, and packets are routed based on the path ID in their headers.
With XPath's explicit path control, \sys adds path IDs to the headers of the encoded symbols, which will place them on different paths.

\section{Evaluation}\label{sec:eval}\label{sec:eval:setting}
In this section, we evaluate \sys with testbed experiments ($\S$\ref{sec:basicdesign}--$\S$\ref{sec:deepdive}), complemented by large-scale simulations ($\S$\ref{sec:simulations}).
We summarize the results below:
\begin{itemize}
\item $\S$\ref{sec:basicdesign}: we inspect \sys's design choices and quantify their benefits. 
With all choices combined, \sys reduces the tail latency by $75.32$\% compared to DCTCP.
\item $\S$\ref{sec:eval:latency}: we compare \sys with the prior practical schemes, and find that it achieves more than $60.06\%$ reduction in p99 flow completion time compared to DCTCP+ECMP~\cite{dctcp,ecmp} or PIAS~\cite{pias-nsdi}.
\item $\S$\ref{sec:deepdive}: we dive into \sys and find that it is resilient to many critical cases including incast and failures.
\item $\S$\ref{sec:simulations}: we use large-scale simulations to show that \sys achieves $24\%$ tail latency reduction compared to a near-optimal clean-slate design, pFabric~\cite{pfabric}.
\end{itemize}

\begin{figure}[t]
  \centering
  \vspace{-1em}
  \includegraphics[width=\linewidth]{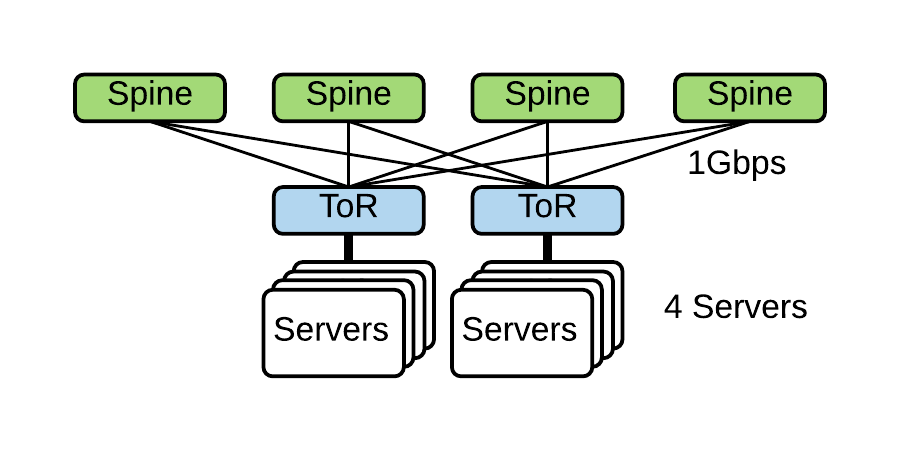}\\
  \vspace{-2em}
  \caption{Testbed setup}\label{fig:testbed}
  \vspace{-1em}
\end{figure}

\textbf{Traffic patterns:} Following related works~\cite{detail}, we emulate traffic of latency-sensitive applications: data retrieval (request/response) and page generation.
Each response message is triggered by a 1.5K-byte (MTU) request from the receiver to the senders. The senders reply with a message with variable size uniformly chosen from $\{5, 10, 20, 50, 93\}$KB.
The inter-arrival time of initiating requests follows exponential distribution with mean $4,5,10$ms (A Poisson random process with arrival rate $250,200,100$ requests per server per second).

Sender-receiver selection is as follows:
\begin{itemize}
\item \textit{All-to-all request/response (by default):} Each server randomly picks another server to send request.
\item \textit{Front/Back-end page generation:} The two racks are designated as Front-end and Back-end. Front-end servers request data from a randomly selected Back-end server.
\end{itemize}

\textbf{Testbed:} We built a spine-leaf (8 servers, 2 leaf or Top-of-rack \& 4 spine switches) testbed to create 4 paths between any pair of servers from 2 racks (Figure~\ref{fig:testbed}). Each path corresponds to a spine switch. We use Pronto-3295 switches and Dell PowerEdge R320 servers, each with a quad core Xeons E5-1410 CPU and 1GbE NIC, and with Debian 6.0 (kernel 2.6.32-5) installed. XPath~\cite{xpath} is enabled by default.

We generate background flows to create network congestion. 
For the 4 servers on one rack, each randomly chooses another server in the other rack, and sends a flow of 10MB using DCTCP on a random path. When a flow finishes, the server will start another one. 
In this way, each path has the same probability for different degrees of congestion ($1/256, 3/64, 27/128, 27/64, 81/256$ chance to have $0, 1, 2, 3, 4$ flows, respectively). 
Unless specified otherwise, background flows are in a separate queue (totally 2 queues are used), and switches use WRR for these 2 queues.

\subsection{Inspecting Design Choices}\label{sec:basicdesign}

\sys incorporates three design choices:
1) \sys uses LTC to encode the message, and runs the "burst-until-received" protocol; 2) \sys spreads the encoded packets on multiple paths, obliviously taking advantage of uncongested paths; 3) the switches perform aggressive dropping with tiny queues for \sys flows. We now study the impact of each of these decisions progressively.

\begin{figure}[t]
  \centering
  \includegraphics[width=\linewidth]{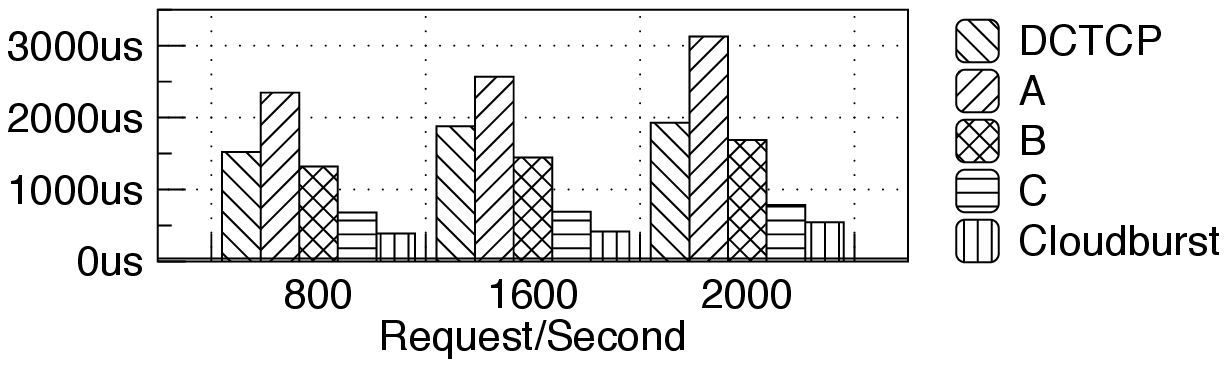}\\
  \vspace{-1em}
  \caption{p99 Completion Time (5KB)}\label{fig:pro:5}

  \includegraphics[width=\linewidth]{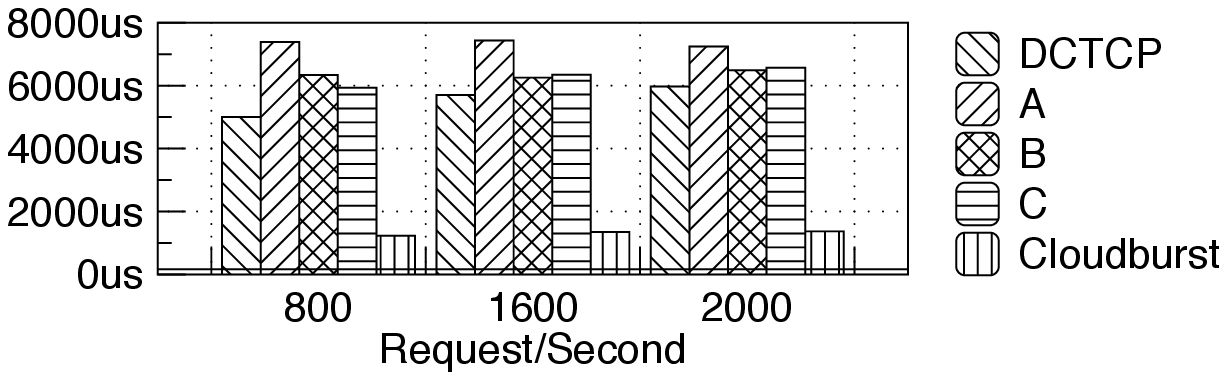}\\
  \vspace{-1em}
  \caption{p99 Completion Time (20KB)}\label{fig:pro:20}

  \includegraphics[width=\linewidth]{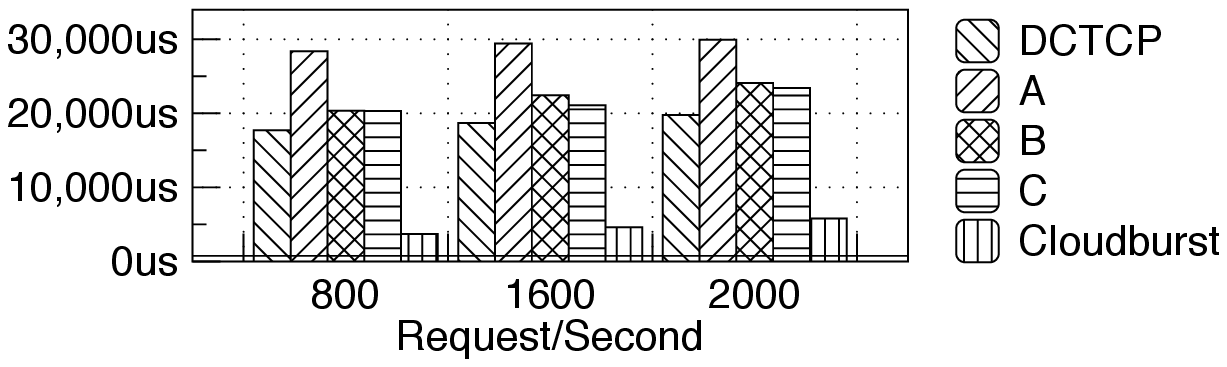}\\
  \vspace{-1em}
  \caption{p99 Completion Time (93KB)}\label{fig:pro:93}
  \vspace{-1em}
\end{figure}

We run the all-to-all pattern with varying request rates (each for 10 minutes), and plot the p99 completion times for $\{5,20,93\}$KB flows in Figure~\ref{fig:pro:5},\ref{fig:pro:20}\&\ref{fig:pro:93}, respectively. 
We take DCTCP as a reference (parameter setting follows $\S$\ref{sec:eval:latency}). We compare the following schemes:
\begin{itemize}
\item \textbf{A}: FEC (Design Choice 1). We encode the message into encoded packets, and send the packets at line rate using UDP on a randomly chosen single path.
\item \textbf{B}: \textbf{A} + Multipath (Design Choice 1\&2). Senders spread the encoded packets of each flow on multiple paths.
\item \textbf{C}: \textbf{A} + Aggressive Dropping (Design Choice 1\&3). Senders send encoded packets on a single path, and switches aggressively drop packets by limiting buffering.
\item \textbf{D}: \sys (Design Choice 1,2,\&3). \sys combines all three design choices.
\end{itemize}

\textbf{Impact of FEC:} 
\sys uses FEC to perform loss recovery. Consider a simple mathematical model: message size is $M$ packets, and link capacity is $C$ packet/s. Assume the packet drop probability is $p_d$ on a path. If a packet is lost, a sender transmitting at link capacity without coding takes $\frac{M}{C}$s to retransmit it, and the expected latency is $E_D=(1-p_d)\frac{M}{C}\sum_{i=0}^{\infty} (i+1)p_d^i =\frac{M}{(1-p_d)C}$. Thus, with a larger $p_d$ (aggressive dropping), $D$ takes more time to recover a lost packet. Encoding essentially sends multiple packets' information at the same time. For degree $d$, $d$ original packets are XOR'd for each encoded packet. In this way, a packet do not have to wait for $\frac{M}{C}$ to be retransmitted. For each encoded packet, the sender randomly chooses $d$ packets, thus it takes $\frac{M}{dC}$ to deliver all the packet's information. Therefore, the time to receive information of all packets becomes $\frac{M}{d(1-p_d)C}$, $d$ times smaller than $E_D$.
However, FEC alone cannot work. Scheme \textbf{A} keeps generating encoded packets, and sends them on a single path. We see that, the tail latency is $51.12\%$ longer compared to DCTCP across all message sizes at $2000$ Request per second (rps). This is caused by the queueing created by the aggressive sending of encoded packets.
In contrast, DCTCP controls the queueing by enforcing a small queue at the switch, which allows the end-hosts to react to congestion quickly.

\textbf{Impact of Multipath:} Queueing is created when background flows choose the same path due to imperfect load balancing. 
Good load balancing is difficult to achieve for TCP flows, because in-order packet delivery is expected. 
However, the same is not true for erasure-coded flows, as there is no order between encoded packets and the original message is reconstructed as a whole. This allows for easy implementation of load balancing: the sender can simply spray encoded packets on multiple paths. Scheme \textbf{B} does exactly this. We observe that, with evenly load balanced FEC flows, the tail latency is improved by $44.52\%$ for $5$KB messages across all loads. This improvement is smaller with increasing message size ($23.88\%$ for $93$KB message), because larger messages take longer to decode (Figure~\ref{fig:coding-overhead}). Compared to DCTCP, we see the same trend: \textbf{B} outperforms DCTCP for short messages (tail latency is $16.24\%$ shorter for $5$KB) and sightly worse for long messages (tail latency is $5.89\%$ longer for $93$KB). This is because Scheme \textbf{B} spreads traffic to all available paths evenly, while DCTCP is vulnerable to imperfect load balancing.

\textbf{Impact of Aggressive Dropping:}\label{sec:eval:drop}
To counter \textbf{A}'s queueing, we can also limit the queue depth. Scheme \textbf{C} uses FEC and aggressive dropping on a single path. In Figure~\ref{fig:pro:20}\&\ref{fig:pro:93}, \textbf{C} is similar to DCTCP for $20$KB and $93$KB messages. For $5$KB short messages, \textbf{C}'s tail latency is $51.42\%$ shorter than that of DCTCP, because DCTCP relies on timeouts to discover packet loss for short messages. In contrast, \textbf{C} proactively retransmits encoded packets despite dropping.

\textbf{Summary:} With all three design points, we have \sys. With erasure coding (Design Choice 1), each encoded packet can help recover any of the $d$ original packets. With multipath forwarding (Design Choice 2), packets are spread on all paths evenly, thus exploiting uncongested paths obliviously. Finally, aggressive dropping (Design Choice 3) ensures that packets arriving at the receiver experience deterministically low queueing delay, because packets that encounter any queue build-up are dropped. Compared to DCTCP, \sys reduces the tail latency by $75.32\%$ (averaged over all sizes).

\subsection{Comparing with prior schemes}\label{sec:eval:latency}
We proceed to compare \sys with existing schemes that are implementable in commodity DCNs:

\begin{itemize}
\item \textbf{DCTCP~\cite{dctcp} + ECMP:} RTO$_{min}$ is set to 10ms. ECN marking threshold is set to 65 packets. All flows share the same switch queue.
\item \textbf{PIAS~\cite{pias-nsdi}:} We implemented PIAS with 2-level feedback queue and set the first threshold to be 95KB, so that all short flows or messages have highest priority. 
\item \textbf{Replicated DCTCP:} Transmitting flows with same content using DCTCP on 2-4 paths~\cite{redundancy} (paths are randomly chosen). All flows share the same switch queue.
\item \textbf{MPTCP~\cite{mptcp,mptcp-dcn}:} Using tc in Linux, MPTCP flows are tagged with DSCP value for the short flow queue.
\end{itemize}

We run the all-to-all request/response traffic pattern, and collect message completion time (MCT) for each scheme.

\begin{figure}[t]
  \centering
  \vspace{-0.5em}
  \includegraphics[width=\linewidth]{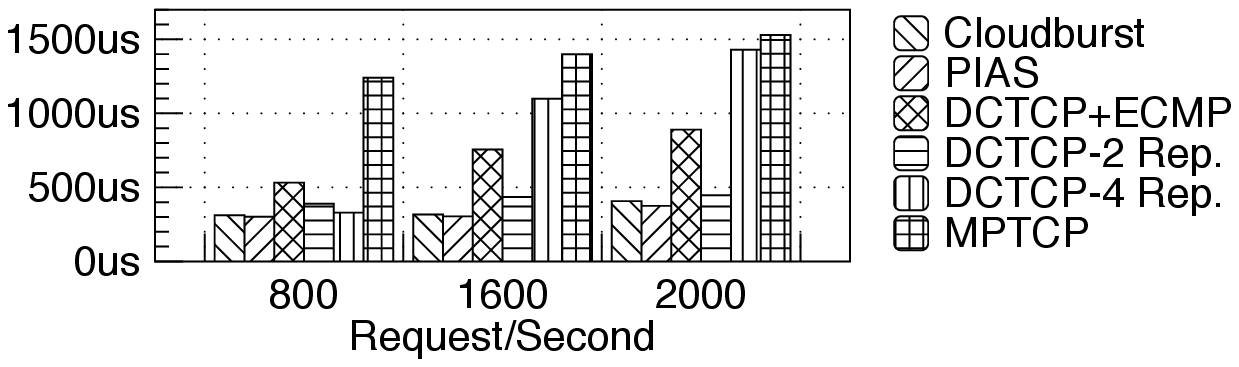}\\
  \vspace{-1em}
  \caption{Average Message Completion Time}\label{fig:eval:avg}

  \includegraphics[width=\linewidth]{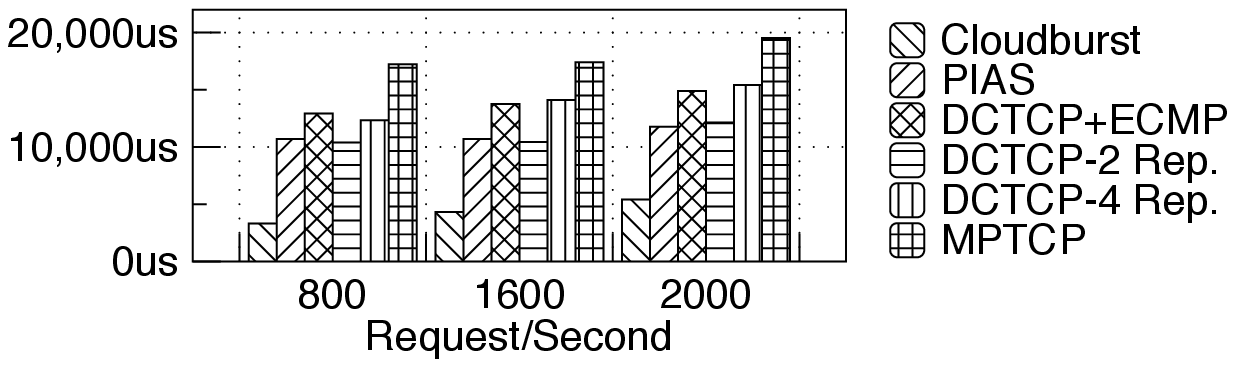}\\
    \vspace{-1em}
  \caption{p99 Message Completion Time}\label{fig:eval:99}

  \vspace{1em}
  \resizebox{0.8\linewidth}{!}{
	\centering
  	\begin{tabular}{l|lll|l}
   		 	& 800 r/s & 1600 r/s 	& 2000 r/s & Average  \\ \hline
    5KB 		& 1.221 	& 1.241   	& 1.381 & 1.281\\
    10KB 		& 1.478 	& 1.133   	& 1.367 & 1.331\\
    20KB 		& 1.398 	& 1.972   	& 1.875 & 1.748\\
    50KB 		& 1.167 	& 1.311  	& 1.643 & 1.373\\
    93KB 		& 1.542 	& 1.928   	& 1.676 & 1.715\\ \hline
    Average		& 1.361 & 1.517		& 1.591	& 1.490
  	\end{tabular}
  }
  \caption{Coding rates of experiments in Figure~\ref{fig:eval:avg},\ref{fig:eval:99}}\label{eval:table}
  \vspace{-1.5em}
\end{figure}

\textbf{Average Latency:} In Figure~\ref{fig:eval:avg}, despite the en/decoding overheads, \sys performs similarly to PIAS for different request arrival rates in terms of average MCT. MPTCP shows the worst performance, because if any congested path exists, MPTCP is bound to experience congestion, prolonging MCT. Among the DCTCP-based schemes, for low request rate (800 r/s), DCTCP with the most duplicated flows (4) achieves the best performance, as it transmits on all paths, thus can always avoid congestion. However, as the request rate increases, the performance of Replicated DCTCP starts to degrade, because the replication essentially multiplies the load, leading to congestion and packet drops.

\textbf{Tail Latency:} In Figure~\ref{fig:eval:99}, for p99 MCT, \sys outperforms all the other schemes. The p99 MCT captures the tail latency events (e.g. long queueing, packet loss). At $2000$ r/s, its tail latency is over $60.06\%$ ($63.69\%$) less than that of PIAS (DCTCP$+$ECMP). This is because \sys packets are encoded with redundancy, thus flows need not wait for retransmission timeout, unlike TCP variants.

\textbf{Coding Overhead:} In Table~\ref{eval:table}, we list the coding rates in Figure~\ref{fig:eval:avg}\&\ref{fig:eval:99}, which is the ratio of the number of sent packets over the number of packets in the original message. The average coding rate is $1.49$, i.e. \sys adds $\sim 49\%$ more traffic load in the experiments.

\subsection{\sys Deep Dive}\label{sec:deepdive}
In this section, we use a series of targeted experiments to further understand \sys.

\begin{figure}[t]
  \centering
  \includegraphics[width=0.95\linewidth]{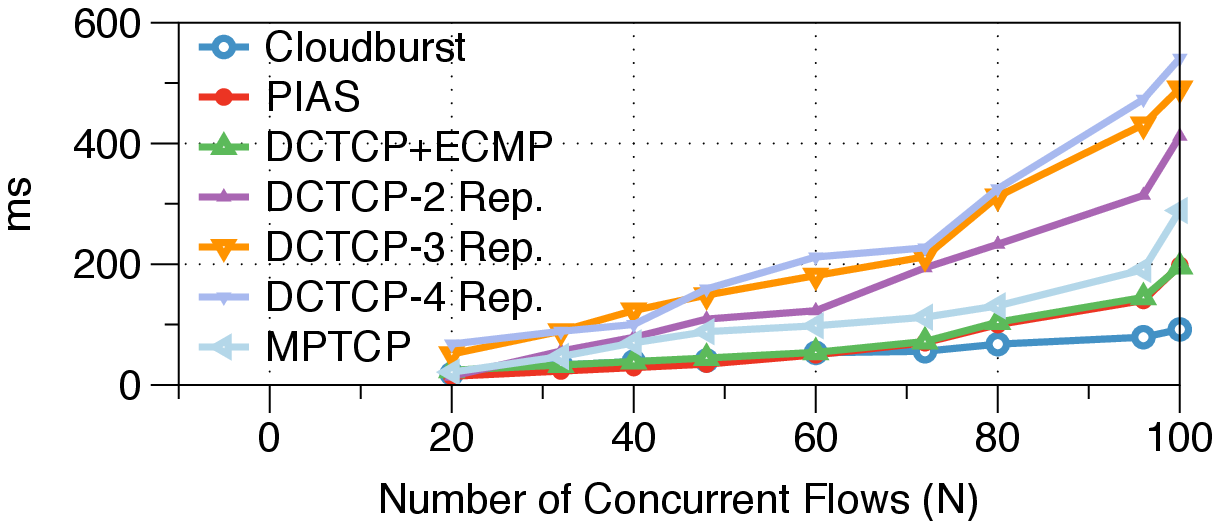}\\
  \vspace{-1em}
  \caption{Incast: Average Completion Time}\label{fig:eval:incast}

  \vspace{0.5em} 
  \includegraphics[width=\linewidth]{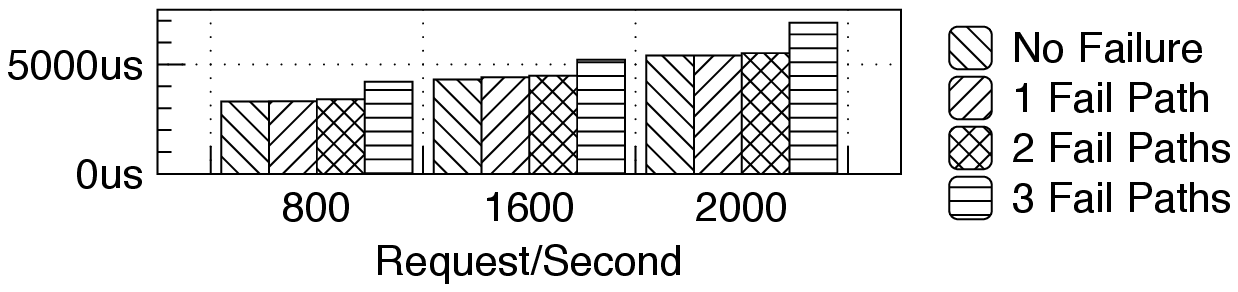}\\
  \vspace{-1em} 
  \caption{Failure: p99 Completion Time}\label{fig:eval:imp}
  \vspace{-1em}
\end{figure}

\textbf{Impact of incast:}\label{sec:eval:incast}
We first examine the incast scenario. We have 4 senders on one rack sending 90KB (60 pkts) messages to a receiver on the other rack, and set receiver's ToR switch buffer size to 100KB (with no traffic to other ports, the total switch buffer size for this port is the sum of shared and dedicated buffer). The flows start at the same time and are evenly distributed among the senders. We increase the number of concurrent flows, $N$, and measure the time from the start to the last flow. 
Figure~\ref{fig:eval:incast} shows the average flow completion times (FCT) with the increase of $N$. We find that, as $N$ grows larger, the performances of different schemes start to diverge, and DCTCP-based schemes (including PIAS) start to have increasingly longer FCT. In contrast, \sys shows consistently low FCT under incast, and its FCT grows almost linearly with $N$. The key reason is that, unlike TCP, \sys's aggressive burst-until-received protocol does not require a timeout to discover packet loss, which is bound to happen in incast. A \sys flow in an incast proactively retransmits without need to discover a packet loss.

\textbf{Impact of failures:}\label{sec:eval:multipath}
When link/switch failures happen, some paths may become unavailable.
We evaluate how \sys handles such scenario. We vary the number of failed paths, and compare p99 FCT.
As shown in Figure~\ref{fig:eval:imp}, with increasing number of failed paths, the p99 FCT increases gradually: for $2000$rps, from $5412\mu$s to $6907\mu$s. As expected, the scenario with the most failed paths performs the worst. This is because, when there is only one path, the tiny buffer on the switch may drop many packets on this path, as \sys senders continuously resend encoded packets on this path to recover the drops. This shows \sys is sensitive to the existence of multipath, but not the number of available paths.

\textbf{Impact to other traffic:}\label{sec:eval:friend}
In this experiment, we assess the impact of \sys flows to other traffic. 
We run the Front/Back-end traffic pattern with \sys for 5 minutes. We initiate 4 DCTCP background flows from Back-end rack to Front-end rack on the 4 paths using iperf, each flow between a different pair of servers. We then increase the bandwidth demand of the \sys flows from $1.424$Mbps per server to $284.4$Mbps (by increasing the request arrival rate from $100$rps to $20000$rps
; For arrival rate $\geq 1000$rps, each request initiate $i$ responses, $i=1,4,7,10,15,18,20$.). On the switch, we configure WRR with weight 4 for background traffic, and 1 for \sys traffic, so that the bandwidth limit for \sys flows is $200$Mbps, smaller than the largest demands. We set the transmission timeout for \sys to be $10$ms (the sender aborts the flow if it is not completed before timeout). We examine the average throughput of background and \sys flows.

\begin{figure}[t]
  \centering
  \includegraphics[width=\linewidth]{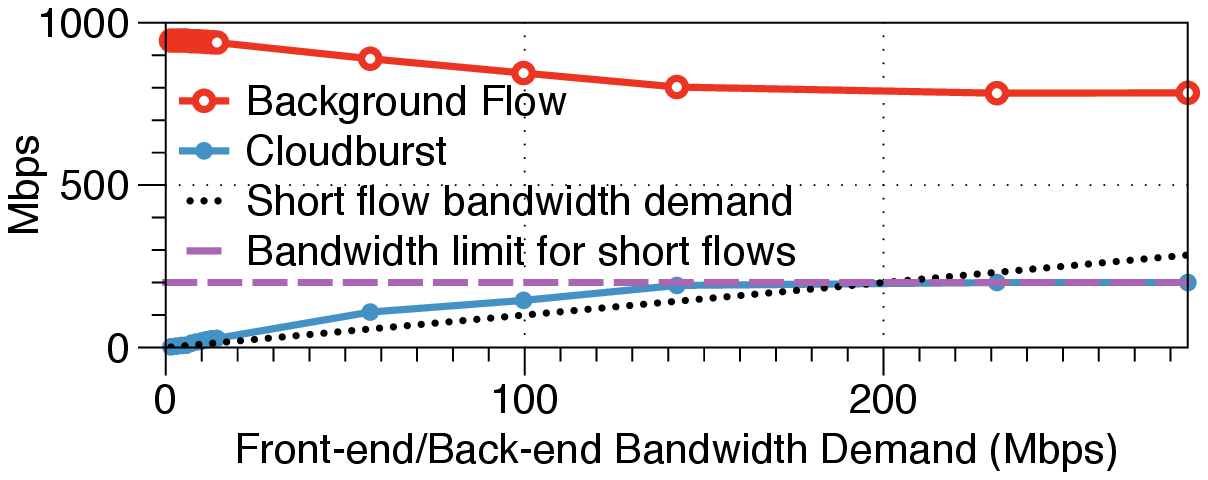}
  \vspace{-2em}
  \caption{Friendliness: background flow throughput}\label{fig:eval:friend}
  
  \vspace{2em}
  
\resizebox{\linewidth}{!}{
\centering
  \begin{tabular}{l|l|l|l|l|l}
    Demand (Mbps) & 99.68 & 142.4 & 213.6 & 256.32 & 284.8 \\ \hline
    Completion \% & 100   & 100   & 43.11 & 29.24  & 12.83 \\
  \end{tabular}
}\caption{Completion Percentages in Figure~\ref{fig:eval:friend}}\label{eval:friend:table}
\end{figure}

Figure~\ref{fig:eval:friend} shows that \sys shares the bandwidth with DCTCP flows. The dotted line shows the demand of the traffic pattern, and the purple dash line is the bandwidth limit for short flows. We see that DCTCP flows are able to maintain high throughput in the presence of \sys flows, owing to the minimal buffer usage of \sys as well as the guaranteed bandwidth from WRR in the switch.

\begin{figure}[t]
  \centering
  \includegraphics[width=\linewidth]{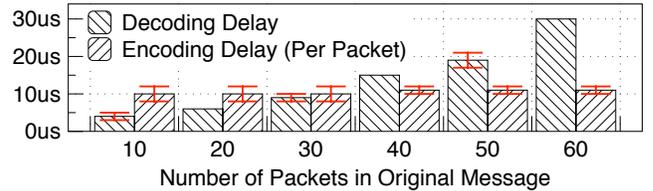}\\
  \caption{En/decoding time}\label{fig:eval:encdec}
\end{figure}

\textbf{Encoding/decoding complexity:}\label{sec:eval:overhead}
We measure the en/decoding latencies for varying message sizes by running \sys on network without background flows.
In Figure~\ref{fig:eval:encdec}, we show that, with increasing number of packets, the encoding overhead remains constant, while the decoding time increases, which indicates that \sys has most to gain by reducing decoding complexity.

\subsection{Large-scale Simulations}\label{sec:simulations}
\begin{figure}[t]
  \centering
  \vspace{-0.5em}
  \includegraphics[width=\linewidth]{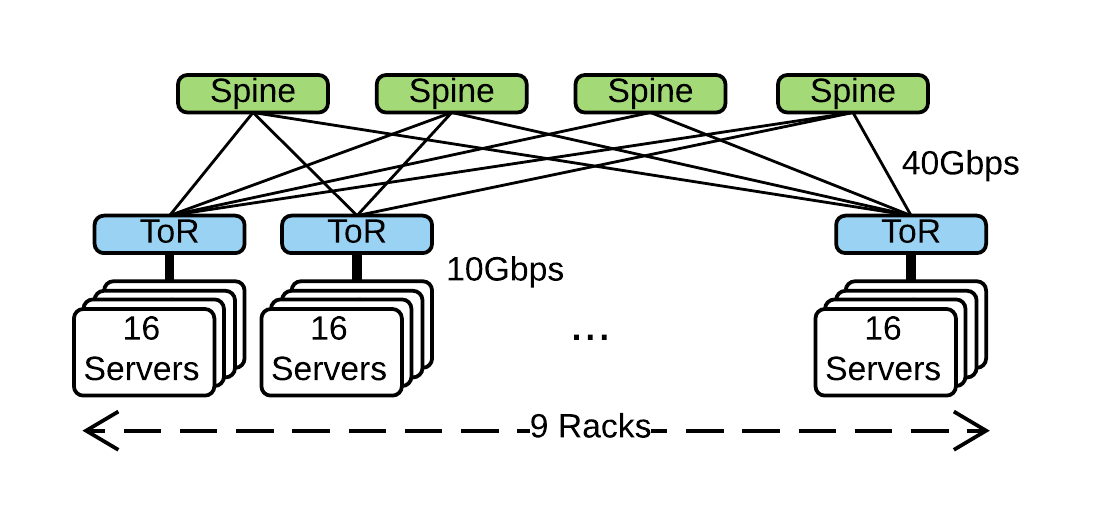}\\
  \vspace{-2em}
  \caption{Simulation Topology}\label{fig:eval:simtopo}
  \vspace{-1em}
\end{figure}

We complement the testbed experiments by simulating \sys in a large DCN (Figure~\ref{fig:eval:simtopo}). We use a leaf-spine topology with 144 hosts, 9 leaf (ToR) switches, and 4 spine (Core) switches. Each leaf switch has $16\times$ 10Gbps downlinks and $4\times$ 40Gbps uplinks to the spine. The base RTT across the spine (4 hops) is 20$\mu$s. We generate the all-to-all traffic workload as above, and fix the traffic load to $2000$rps.
We implemented \sys based on ns-2 simulation of QJump~\cite{qjumpns2}. The en/decoding times of \sys are obtained from experiments (not shown due to space limitation), and are added to \sys's FCT measurement. We compare \sys with the following:
\begin{itemize}
  \item \textbf{DCTCP~\cite{dctcp}:} We configure DCTCP with the recommended ECN marking threshold of 65 packets.
  \item \textbf{pFabric~\cite{pfabric}:} Queue size is 34 packets ($2\times$BDP), initial window is 17 packets, and RTO$_{min}$ is 1ms.
  \item \textbf{QJump~\cite{qjump}:} Based on topology, we configure the minimum bandwidth $R=10$Gbps, cumulative switching delay $\epsilon=4$us, $P=MTU=1.5$KB. For messages, we set the throughput factor $f=1$ (guaranteed latency); for background flows, we set $f=n$ (maximum throughput. $n=144$, the total number of end-hosts).
  \item \textbf{Expresspass\cite{expresspass}:} For credit packet, the packet size is 84 bytes and the queue size is 10 packets; for data packet, the packet size is 1538 bytes and the queue size is 100 packets. The credit rate is 500Mbps.
\end{itemize}

\begin{figure}[t]
  \centering

  \includegraphics[width=\linewidth]{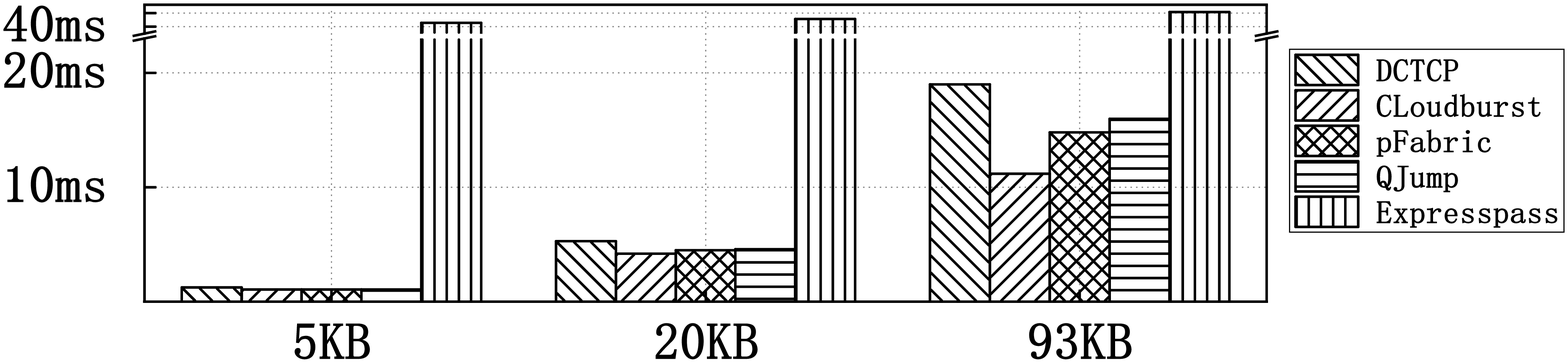}\\
  \vspace{-1em}
  \caption{p99 Completion Time}\label{fig:eval:over}

  \vspace{1em}
  
  \includegraphics[width=\linewidth]{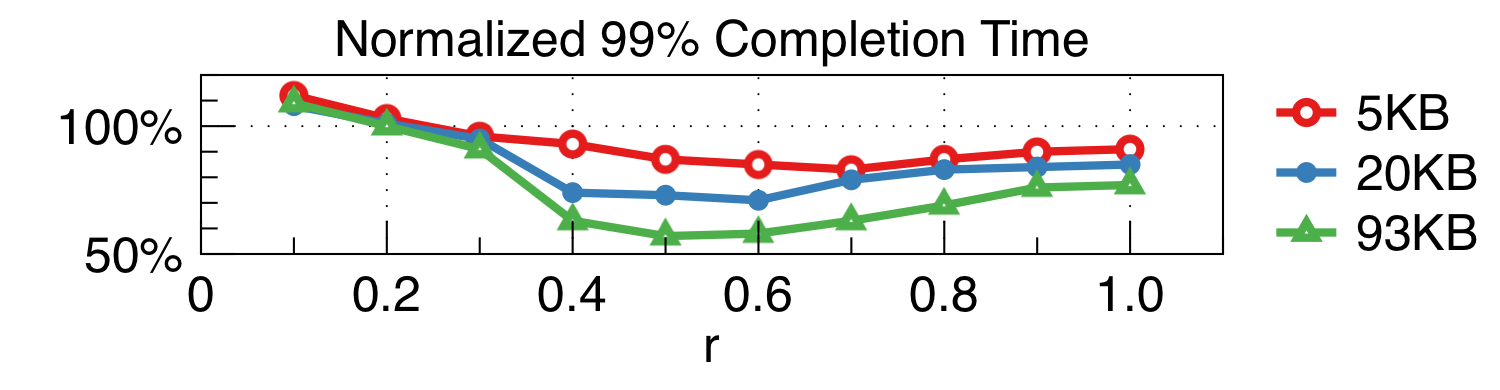}\\
  \vspace{-1em}
  \caption{Choosing $r$ in 2:1 over-subscribed network}\label{fig:eval:r-over}
  
  \vspace{1em}
  
  \includegraphics[width=\linewidth]{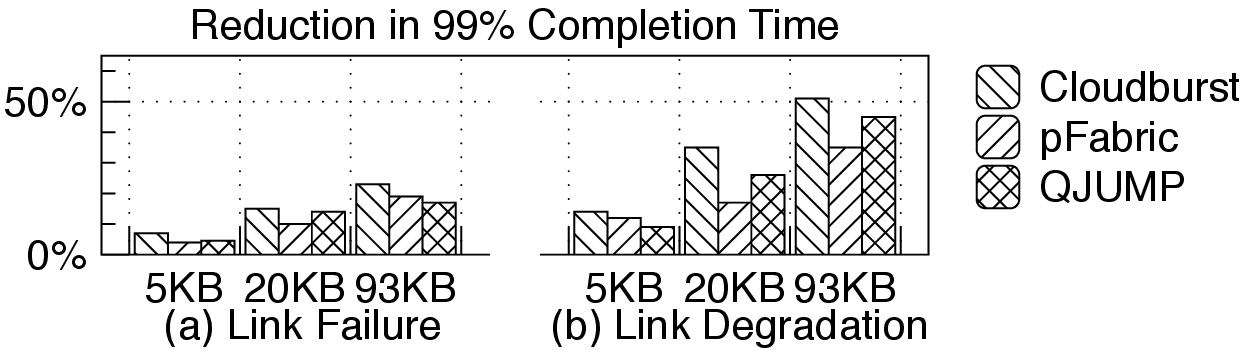}\\
  \vspace{-1em}
  \caption{Two failure scenarios}\label{fig:eval:fail}
  
\end{figure}

\textbf{Tail latency:}
We first compare the tail latency. Figure~\ref{fig:eval:over} shows p99 FCT. For small message size ($5$KB), the difference is insignificant. As size grows larger, \sys begins to show its advantage over the others. The main reason is that TCP-like schemes takes at least RTO$_{min}$ to recover loss. Since packet loss is captured by the tail latency, \sys outperforms DCTCP, pFabric, and QJump by $40.12\%$, $24\%$, and $29.63\%$, respectively. To our surprise, our result shows that Expresspass performs the worst among all the schemes. We imagine there are two main reasons: 1) it requires one additional RTT for credit allocation, and 2) such credit-based algorithm is fundamentally not suitable for mice flows in a dynamically changing environment, because it is very hard to set up an appropriate amount of credits for short-live flows.

\textbf{Impact of over-subscription:}\label{sec:eval:oversub}
We further examine how \sys performs in over-subscribed networks. We reduce all ToR-to-Core links' capacity to 20Gbps, creating a network with 2:1 over-subscription. We run this experiment for DCTCP and \sys. For \sys,
we vary $r$, the rate limit in Algorithm~\ref{algo:send}, from $0.1$ (sending at $1$Gbps) to $1$ ($10$Gbps), and collect MCTs.
We summarize the results in Figure~\ref{fig:eval:r-over} normalized to DCTCP.
We find that, for varying sizes, choosing $r$ close to the over-subscription ratio results in higher tail latency reduction. If $r$ is too small, the sender is not sending enough to compensate for the aggressive dropping, and the message can take longer to finish. If $r$ is too large, the senders may overload the network and cause more frequent packet drops, which is bad for longer messages. 
Overall, when $r$ is chosen appropriately ($\sim0.5$), the MCT reduction is $>12.88\%$ for all message sizes. 
Therefore, we suggest setting the sending rate $r$ (in Algorithm~\ref{algo:send}) to the over-subscription ratio.

\textbf{Impact of failures:}\label{sec:eval:fail}
Finally, we examine two failure scenarios at large scale.
We run all-to-all pattern. For link failure scenarios, we randomly remove one of the Core-ToR links.
For link degradation scenarios, we randomly decrease one Core-ToR link's capacity to 1Gbps.
Figure~\ref{fig:eval:fail} shows both scenarios, and the metric is also reduction percentage compared to DCTCP.
We observe that, for both scenarios, \sys is tolerant of failures, and can achieve reduction similar to that in Figure~\ref{fig:eval:over}. This is due to \sys's inherent randomness. As discussed in $\S$\ref{sec:eval:multipath}, \sys benefits from multipath, but is not sensitive to the number of available paths.

\section{Discussion}\label{sec:discuss}
In this section, we address existing issues, and discuss concerns and further improvement of \sys.



\textbf{Hardware en/decoding:} The overhead of \sys can be further reduced if the en/decoding is offloaded to hardware (e.g., programmable NIC). The encoding is constant time (given degree $d$), but the decoding using Gaussian elimination has $O(k^2)$ time complexity ($k$ is the number of packets in original message). Although $k$ is always small for \sys, offloading coding to dedicated hardware still significantly reduces the stress on end-hosts. In future high-speed networks, hardware offload is inevitable, because the computational load to en/decode at 100/400Gbps line rate will overburden the CPUs. We leave design of such hardware as future work.


\textbf{Using other rateless codes:}
Coding scheme is not closely coupled with \sys, and other rateless codes can work as well (with modification to the header format in Figure~\ref{fig:header}).
LTC is chosen for its simplicity, so that we can quickly validate our idea of coding over multipath.
We note that more efficient coding schemes, such as Raptor codes~\cite{raptor} which have linear en/decoding complexity, are good alternatives.


\textbf{Fairness among \sys flows:} \sys achieves fairness with aggressive dropping with tiny buffers, because dropping is a form of congestion control~\cite{decongestion} for erasure-coded flows. It is shown that, if congested link drop packets in a fair manner~\cite{diffdrop}, each flow will receive its max-min fair throughput. For a network with purely erasure coded flows, the switches do not need deep buffers to keep network stable~\cite{nocc}, and the sources can greedily send as fast as possible~\cite{decongestion}.


\textbf{\sys for long flows:}
Coding mechanisms provide reliability at the cost of redundancy, so we carefully limit \sys to short messages. For decoding to work, the number of packets in the original message should be received first. The bit-set indicating which of the original packets are included in the received encoded packet is also necessary for the decoding process. We put all the necessary in the header (Figure~\ref{fig:header}), so the decoding process can start no matter which encoded packet arrives first. The header overhead increases with the number of packets in a message. Since the header size must be less than MTU, the MTU of the network therefore limits the number of packets in a message.
Using rateless codes for long flows, the coding rate can be unbounded, and the resulting redundancy can overload the network fabric. Consider the situation when long flows using \sys collide with each other on multiple paths. The long flow will repeatedly send out redundancy packets trying to overcome the poor channel condition, which worsens congestion gravely.
Splitting a long flow into several small \sys flows is a work-around. In this way, a long flow is segmented into a series of short flows, and \sys delivers each segment.

\textbf{\sys with programmable data plane:}
With advanced data plane, we envision the following improvements to \sys: (1) \textit{True zero queueing delay} for \sys packet can be achieved by dropping incoming \sys packets whenever a \sys packet is in buffer. Current buffer limit (2\%) is needed to maintain full port throughput, as setting it to 0 drops all packets.
(2) Instead of dropping based on each switch's local queue limit, we can make \textit{per-packet dropping decisions}, because packets themselves can accumulate the queueing latency values over multiple hops and let intermediate switches drop packets when the cumulative queueing latency (i.e., experienced path latency) exceeds a threshold.
Per-packet latency can be recorded in a field in the packet header, and the packet can be dropped if its experienced delay exceeds a threshold.
Dropping these packets will reduce the load on later switches on the path, and give more opportunity for other packets to go through.



\section{Related work}\label{sec:related}
In addition to prior solutions ($\S$\ref{sec:background:prior}),
we briefly survey related work in applications of coding on multipaths.


Coding techniques over multipath are extensively employed to improve transport layer in various settings, including wireless environments~\cite{huang2008tcp, sundararajan2009network, hmtp}, vehicular networks~\cite{oec}, and information dissemination in overlay networks~\cite{bullet}.
They use FEC to mitigate impact of packet loss, and reduce retransmission. A similar work that employs fountain coding over multipath is HMTP~\cite{hmtp}, which solves the receiver packet reordering problem in multi-homing wireless networks.
Another example of FEC over multipath is OEC in vehicular networks~\cite{oec}, which has a streaming (receiver-feedback) setting and a greedy encoding scheme to maximize new information at the receiver.
Relative to them, \sys applies FEC over multipath for DCN tail latency reduction.

More recently, FMTCP~\cite{fmtcp} combines fountain code with MPTCP~\cite{mptcp}, in order to mitigate MPTCP's problem of being straggled by the slowest subflow. FMTCP employs a data allocation algorithm to coordinate the coding and transfer of many subflows. 
QUIC~\cite{quic} also uses FEC to mask packet drops in order to reduce web latency. Specifically, it uses proactive speculative retransmission (sending duplicated copies of more important packets) to optimize streaming applications. In contrast to them, \sys employs aggressive dropping to reduce per-hop delay, and focuses on reducing delivery latency of short flows in DCNs.


\section{Conclusion}\label{sec:conclusion} 

We present the design, implementation, and evaluation of \sys ~--- a simple scheme to cut long tail latency of message delivery by proactively sending FEC-coded packets generated from the messages on multipath in parallel, thus obliviously exploiting the uncongested paths without complexities like prior solutions. \sys is readily deployable in today's commodity DCNs. We implemented a \sys prototype, and validated its performance with extensive testbed experiments as well as large-scale simulations.


\end{document}